\documentclass[]{tPHM2e}
\usepackage{amsfonts}
\usepackage{amssymb}
\usepackage{graphicx}
\usepackage{color}
\begin{document}

%\issn{1478-6443}
%\issnp{1478-6435}
%\jvol{00} \jnum{00} \jyear{2010} %\jmonth{21 December}

\markboth{Covariant Hysteretic Constitutive Theory for Maxwell's equations}{}

\title{{\itshape Covariant Hysteretic Constitutive Theory for Maxwell's equations}: Application to Axially Rotating Media}

\author{Alison C Hale$^{\rm a}$$^{\rm b}$$^{\ast}$\thanks{$^\ast$Corresponding author. Email: a.c.hale@lancaster.ac.uk
\vspace{6pt}} and Robin W Tucker$^{\rm a}$$^{\rm b}$\\\vspace{6pt}  $^{\rm a}${\em{Department of Physics, Lancaster University, Lancaster, LA1 4YB, UK}}; $^{\rm b}${\em{Cockcroft Institute of Accelerator Science and Technology, Daresbury, Keckwick Lane, Daresbury,  WA4 4AD, UK}}\\
\vspace{6pt} }

\maketitle

\begin{abstract}
This paper explores a class of non-linear constitutive relations for materials with memory in the framework of covariant macroscopic Maxwell theory. Based on earlier models for the response of hysteretic ferromagnetic materials to prescribed slowly varying magnetic background fields, generalized models are explored that are applicable to accelerating hysteretic magneto-electric  substances coupled self-consistently to Maxwell fields. Using a parameterized model consistent with experimental data for a particular material that exhibits purely ferroelectric hysteresis when at rest in a slowly varying electric field, a constitutive model is constructed that permits a numerical  analysis of its response to a driven harmonic electromagnetic field in a rectangular cavity. This response is then contrasted with its predicted response when set in uniform rotary motion in the cavity.
\end{abstract}

\bigskip

%77.80.Dj, 75.60.-d, 77.80.-e, 75.78.-n, 75.85.+t, 41.20.-q, 03.50.De, 02.40.-k
%Hysteresis in ferroelectricity, 77.80.Dj
%Hysteresis in magnetism, 75.60.-d
%Ferroelectricity, 77.80.-e
%Magnetization, dynamics of, 75.78.-n
%Multiferroics, 75.85.+t
%Electromagnetism, 41.20.-q
%Electrodynamics, classical, 03.50.De
%Differential geometry, 02.40.-k

\section{Introduction}

\def\EM{ electromagnetic }
\def\ud{ \underline{d}  }

In recent years developments in technology have enabled the synthesis of new materials with interesting mechanical and electromagnetic properties. These have, in turn, led to new advances in technology as well as challenges in our understanding of physics at the interface between classical and quantum behaviours. In particular many meta-materials on the mesoscopic scale have a rich  electromagnetic  phenomenology. In the presence of high frequency or high intensity electromagnetic fields many exhibit local or non-local  non-linear electromagnetic constitutive properties. At the other extreme some materials exhibit   a delayed response to slowly varying electric or magnetic fields. {\it Ferromagnetic} media respond with a delayed magnetisation while  {\it ferroelectrics} respond with a delayed  electric polarisation. They also exhibit an ability to maintain a saturated internal magnetisation (ferromagnets) or internal electric polarisation (electrets). Anisotropic {\it magneto-electric} materials also exist that sustain both types of fields. When the internally induced fields are not uniquely determined by any externally applied field one often says that the material exhibits  electromagnetic constitutive properties with memory. When the  external fields vary periodically with time  and the induced fields respond periodically,  the process is often referred to as hysteretic and the corresponding
non-linear constitutive relation between them may exhibit a discontinuous or branched structure to account for this  \cite{mil},    \cite{ismail},  \cite{ben}.
However such terminology is often restricted to processes where the induced fields in the medium saturate at some level  and where the {\it shape} of the resulting hysteresis loop, obtained by displaying the magnitude of the drive field against the magnitude of the induced magnetisation or polarisation, is independent of time.
When the drive field is aperiodic in  time the memory effects may exhibit a more complex Lissajoux structure in the hysteretic response, particularly if the time dependent drive field contains more than one dominant harmonic component.

 A rapidly varying time-harmonic electromagnetic drive field may also induce both electric and magnetic polarisations with magnitudes dependent on the magneto-electric susceptibility 3-tensors of the medium. For materials with memory such susceptibilities will depend non-linearly on the electromagnetic field in the medium and the resulting hysteretic response will involve both induced polarisations. In addition to hysteretic responses all materials exhibit spatial and temporal dispersion to some degree and may also sustain induced electric currents as a result of their conductivity. Even in homogeneous and isotropic media the detailed description of such media in terms of their basic constituents and micro-structure is rarely possible and recourse to a parametrised model becomes necessary  \cite{vis}. The parameters of such a phenomenological model are sought from experiment over some range and the model tentatively extrapolated outside that range. The degree of extrapolation is often dictated by comparing the model with experiment.

In this article a model is constructed that  can describe a  rigid non-dispersive,  rapidly, uniformly rotating,  hysteretic medium in an external time-harmonic  electromagnetic field given its behavior at rest.
The model assigned to the medium at rest is motivated by a non-covariant model for magnetic materials constructed by Coleman and Hodgdon \cite{coleman1}, \cite{coleman2}. By exploiting the inherent spacetime covariance of the macroscopic Maxwell equations such a model can be embedded in a covariant formulation and coupled naturally to such equations. This system can then be reduced to a coupled differential system in terms of electromagnetic fields and spatial tensors describing the magnetisation and polarisation in an arbitrarily moving medium. To illustrate how such a system can have practical implications these equations are attacked numerically for an axially rotating  hysteretic  ferroelectric in a perfectly conducting cavity containing  electromagnetic fields driven by an  external harmonic electric current.

\section{Macroscopic Covariant Electrodynamics}

A theory will be said to admit a spacetime covariant formulation if it can be expressed in terms of tensor field equations on spacetime.
The covariant theory of macroscopic electrodynamics benefits from a formulation in terms of differential forms. Aside from its elegance it makes precise a number of conceptual terms used in the interpretation of the theory  of accelerated media and provides a suite of economical tools that streamline calculations. Such tools include the exterior product, exterior derivative, Lie derivative, covariant derivative, interior derivative and Hodge map  \cite{rwt}. These operations find their natural setting on arbitrary manifolds of arbitrary dimension. In this article they are employed on 3-dimensional Euclidean space and 4-dimensional Minkowsi spacetime. In the former case the Hodge map is denoted by $\#$ and satisfies the rule $\#\#=1$ when acting on all (time-dependent) differential forms on space.  The exterior derivative on such forms is denoted $ \underline{d} $  and satisfies $ \ud\ud=0  $. These two identities suffice to determine the many interrelations between the curl operator ($\#\ud$) and div operator ($\#\ud\#$) in Euclidean space. In spacetime the Hodge map is denoted by $\star$ and satisfies $\star\star=(-1)^{1+p}$ when acting on $p-$forms on spacetime. The exterior derivative of such $p-$forms on spacetime is denoted $ d $  and satisfies $ d d =0 $. Although gravitation is neglegible in the following the relations between the frame-dependent spatial description of  electromagnetism and its frame-independent spacetime description and the  Hodge map that enters via the constitutive modelling require a metric tensor field $\mathbf{g} $ for their formulation. To this end one introduces a set of independent cobasis $1-$forms $e^0,e^1,e^2,e^3$ on spacetime and writes the
Minkowski metric tensor field
\begin{equation}
\mathbf{g}=-e^{0}\otimes e^{0}+\underline{\mathbf{g}}
\end{equation}
where $\underline{\mathbf{g}}=\sum\limits_{k=1}^{3}e^{k}\otimes e^{k}$ is the induced metric tensor on an Euclidean space.
In terms of these forms,
 $ \star 1=e^{0}\wedge
e^{1}\wedge e^{2}\wedge e^{3}$,  and
 $\#1=e^{1}\wedge e^{2}\wedge e^{3}$.
The macroscopic Maxwell system on spacetime is defined in terms of the  electromagnetic $2-$form $F$, a polarisation $2-$form $\Pi$  and a $4-$current $3-$form $ \mathcal{J} $:

\begin{eqnarray}
dF &=&0  \label{equ:dsG} \\
d\star G &=&\mathcal{J} \label{equ:dsG2}
\end{eqnarray}
where
\begin{equation}
G=\epsilon _{0}F+\Pi
\end{equation}
in terms of the permittivity of free space $ \epsilon _{0}$.
It is the responsibility of constitutive theory to provide auxiliary equations to render an augmented system deterministic. Thus constitutive auxiliary conditions should specify the dependence of $\Pi $  and  the $4-$current $\mathcal{J}$ (subject to $4-$current conservation,  $ d\mathcal{J} =0 $),  on $F$ and its possible derivatives. Such conditions may be local or non-local, linear or non-linear, algebraic or differential.

 An arbitrary observer in spacetime can be associated with the integral curve of an arbitrary future pointing unit time-like vector field $V$:
 $ \mathbf{g} (V,V)=-1$.
Given $  \mathbf{g}  $,  such a  vector field  determines
the  $1-$form $\widetilde{V}$  by
\begin{equation}
\widetilde{V}=\mathbf{g}(V,-)
\end{equation}
It is convenient to refer to such a $V$ as a frame (in spacetime) since it determines the components of spacetime tensors measured by the associated observer.
Thus the  $2-$forms $F,G,\Pi $ admit the orthogonal
decompositions with respect to $V$:

\begin{eqnarray}
F &=&{\cal E}^{V}\wedge \widetilde{V}-cB^{V}={\cal E}^{V}\wedge \widetilde{V}+\star \left(
c{\cal B}^{V}\wedge \widetilde{V}\right) \\
G &=&{\cal D}^{V}\wedge \widetilde{V}-\frac{H^{V}}{c}={\cal D}^{V}\wedge \widetilde{V}
+\star \left( \frac{{\cal H}^{V}}{c}\wedge \widetilde{V}\right) \\
\Pi &=&{\cal P}^{V}\wedge \widetilde{V}-\frac{M^{V}}{c}={\cal P}^{V}\wedge \widetilde{V}
+\star \left( \frac{{\cal M}^{V}}{c}\wedge \widetilde{V}\right)  \label{equ:PI} \\
\mathcal{J} &=&\frac{-J^{V}}{c}\wedge \widetilde{V}-\rho ^{V}\#1
\end{eqnarray}
where $i_{V}{\cal E}^{V}=0,$ $i_{V}{\cal D}^{V}=0,$ $i_{V}{\cal P}^{V}=0,i_{V}J^{V}=0$ $,$ $ i_{V}B^{V}=0,$
$i_{V}H^{V}=0,$ $i_{V}M^{V}=0,$ $i_{V}( \rho ^{V}\#1 )=0$.
 In these expressions ${\cal E}^V$,  $E^V$, denote a (time-dependent) spatial electric field $1-$form and  $2-$form respectively;
  ${\cal D}^V$,  $D^V$, denote a (time-dependent) spatial electric displacement field $1-$form and  $2-$form respectively;
  ${\cal B}^V$,  $B^V$, denote a (time-dependent) spatial magnetic induction
field $1-$form and  $2-$form respectively;
   ${\cal H}^V$,  $H^V$, denote a (time-dependent) spatial magnetic
field $1-$form and  $2-$form respectively;
   ${\cal P}^V$,  $P^V$, denote a (time-dependent) spatial electric polarisation
field $1-$form and  $2-$form respectively;
   ${\cal M}^V$,  $M^V$, denote a (time-dependent) spatial magnetic polarisation
field $1-$form and  $2-$form respectively;
   $J^V$ denotes a (time-dependent) spatial $3-$current $2-$form and
 $\rho^V$ denotes a (time-dependent) spatial electric charge density  $0-$form.
It should be stressed that these decompositions are defined for arbitrary observer fields, including those describing accelerated frames.
For typographical clarity we will sometimes denote $i_Y \alpha$ by $\alpha (Y)$ for a generic $p-$form $\alpha$.

In Minkowski spacetime there exist global coordinates $t,x,y,z$ in which the above ${\mathbf{g}}-$orthonormal cobasis takes the form:
\begin{equation}
e^{0}=c\,dt,\, e^{1}=dx,\, e^{2}=dy,\, e^{3}=dz  \label{equ:cartbasis}
\end{equation}
The history of an inertial observer is then part of an integral curve of the vector field
\begin{eqnarray}
U &=&\frac{1}{c}\partial _{t}
\end{eqnarray}
and
\begin{eqnarray}
\widetilde{U} &=&-cdt
\end{eqnarray}
In such a spacetime coordinate system the inertial frame time rate of change of any time dependent $p-$form $\alpha$ is defined as  the  Lie derivative  $c{\mathcal L}_U\,\alpha$ and denoted $\dot\alpha  $.

If one makes the orthogonal decompositions above with respect to such an inertial (laboratory) frame the Maxwell system (\ref{equ:dsG}) yields:

\begin{eqnarray}
\#\underline{d}\#{\cal B}^{U} &=&0\quad or \quad \nabla \cdot
\mathbf{B}^{U}=0 \\
\#\underline{d}{\cal E}^{U} &=&-\dot{{\cal B}}^{U}\quad or\quad \nabla
\times \mathbf{E}^{U}=\mathbf{-\dot{B}}^{U}  \label{equ:curlE}
\end{eqnarray}
and (\ref{equ:dsG2}) yields
\begin{eqnarray}
\#\underline{d}\#{\cal D}^{U} &=&\rho ^{U}\quad or \quad \nabla
\cdot \mathbf{D}^{U}=\rho ^{U} \\
\#\underline{d}{\cal H}^{U} &=&\dot{{\cal D}}^{U}+j^{U}\quad or\quad \nabla \times
\mathbf{H}^{U}=\mathbf{-\dot{D}}^{U}+\mathbf{J}^{U} \label{equ:curlH}
\end{eqnarray}
where $j^{U}=\#J^{U}$, $E^U=\#{\cal E}^{U}$, $D^U=\#{\cal D}^{U}$, $B^U=\#{\cal B}^{U}$, $H^U=\#{\cal H}^{U}$ and $\dot{\alpha}=\mathcal{L}_{\partial t}\alpha $ for any $\alpha$. \ The bold-face
characters $\mathbf{E}^{U}, \mathbf{D}^{U}, \mathbf{B}^{U}, \mathbf{H}^{U}, \mathbf{J}^{U}$ refer to
the traditional Gibb's time dependent $3-$vector fields in the inertial (laboratory) frame $U$.
The relation $G=\epsilon _{0}F+\Pi$ yields

\begin{eqnarray}
\mathbf{D}^{U}=\epsilon _{0}\mathbf{E}^{U}+\mathbf{P}^{U} \\
\mathbf{H }^{U}=\frac{1}{\mu_{0}}\mathbf{B}^{U}+\mathbf{M}^{U} \label{equ:Hconstitutive}
\end{eqnarray}
and substituting these into the Maxwell equations (\ref{equ:curlE},\ref {equ:curlH})
gives,
\begin{eqnarray}
\dot{{\cal H}}^{U} &=&-\frac{1}{\mu _{0}}\#\underline{d}{\cal E}^{U}+\dot{{\cal M}}^{U}\quad or \quad \mathbf{\dot{H}}^{U}=-\frac{1}{\mu _{0}}\nabla \times
\mathbf{E}^{U}+\mathbf{\dot{M}}^{U}  \label{equ:Yee1} \\
\dot{{\cal E}}^{U} &=&\frac{1}{\epsilon _{0}}\left( \#\underline{d}{\cal H}^{U}-j^{U}-\dot{\cal P}^{U}\right) \, or\, \mathbf{\dot{E}}^{U}=\frac{1}{ \epsilon _{0}}\left( \nabla \times
\mathbf{H}^{U}-\mathbf{J}^{U}-\mathbf{ \dot{P}}^{U}\right)  \label{equ:Yee2}
\end{eqnarray}

\section{Covariant Constitutive Models}

To set the models to be discussed in context it is worth recalling the simplest covariant constitutive model describing a
non-dispersive, non-hysteretic, homogeneous, isotropic linear material with arbitrary $4-$velocity $V$:
\begin{equation}
G=\epsilon _{0}\left( \epsilon _{r}-\frac{1}{\mu _{r}}\right) i_{V}F\wedge
\widetilde{V}+\frac{\epsilon _{0}}{\mu _{r}}F  \label{equ:mink0}
\end{equation}
where $\epsilon _{r}$ and $ {\mu _{r}}$ are dimensionless constants.
In the co-moving frame $V$ these yield
\begin{equation}
\mathbf{D}^{V}=\epsilon _{0}\epsilon _{r}\mathbf{E}^{V}, \quad
\mathbf{H}^{V}=\frac{1}{\mu _{0}\mu _{r}}\mathbf{B}^{V}
\label{equ:stdcons}
\end{equation}
with $\mu_0=\frac{1}{c^2 \epsilon_0}$.  The electric displacement in the co-moving frame is proportional to the electric field in that frame and the magnetic field in the co-moving frame is proportional to the magnetic induction in that frame.
If $G$ and $F$   in (\ref{equ:mink0})      are decomposed with respect to different frames (possibly accelerating) then the relations between the electric and magnetic fields in different frames  are different and involve the instantaneous relative $3-$velocities between the frames. This is often referred to as a {\it motion  induced} magneto-electric effect.

Since $\Pi =G-\epsilon _{0}F$, $\ \ \star (
( i_{V}\star G) \wedge \widetilde{V}) =-(1+\widetilde{V}\wedge i_{V})G$,   and $
i_{V}G=\epsilon _{0}\epsilon _{r}i_{V}F$ equation (\ref{equ:mink0}) may be written as an algebraic local relation between $\Pi$, $F$, $G$ and  $V$:
\begin{equation}
\Pi =\epsilon _{0}\left( \epsilon _{r}-1\right) i_{V}F\wedge \widetilde{V} +\left( \mu
_{r}-1\right) \star \left( \left( i_{V}\star G\right) \wedge \widetilde{V}\right)
\label{equ:mink1}
\end{equation}

If the medium is conducting a constitutive relation for a conductivity current is required.  In an inertial frame $U$  with local coordinates $t,x,y,z$ suppose the total current density $1-$form is
\begin{equation}
j^{U}(x,y,z,t)=j_{cond}^{U}(x,y,z,t)+j_{ext}^{U}(x,y,z,t)
\end{equation}
where $ j_{ext}^{U}(x,y,z,t) $ denotes a prescribed external current.
A simple isotropic Ohmic conductivity current arises from the temporal non-local relation:

\begin{equation}
j_{cond}^{U}(x,y,z,t)=\int_{-\infty }^{t}\kappa (t^{\prime
}-t){\cal E}^{U}(x,y,z,t^{\prime })dt^{\prime }
\end{equation}
since its Fourier transform with respect to $t$ yields:
\begin{equation}
\hat{j}_{cond}^{U}(x,y,z,\omega^U )=\sigma (\omega^U )\hat{{\cal E}}^{U}(x,y,z,\omega^U )
\end{equation}
where the scalar conductivity $\sigma (\omega^U )$ is the Fourier transform of the spatially homogeneous scalar $\kappa (t)$,
$\hat{j}_{cond}^U$ is the Fourier transform of $j_{cond}^U$
 and
$\hat{{\cal E}}^U$ is the Fourier transform of ${\cal E}^U$.
A spacetime  model therefore requires a model for $\kappa$. In practice one finds data
for $\sigma(\omega^U)$ over some restricted range of $\omega^U$ which is rarely sufficient, in general, to fully re-construct $\kappa$ by Fourier inversion. However in circumstances where the electromagnetic fields have  Fourier components dominantly in the frequency range of relevance the approximation
\begin{equation}
j^{U}_{cond}(x,y,z,t)\simeq \sigma(\omega^U ){\cal E}^{U}(x,y,z,t)
\end{equation}
often suffices. In this approximation $\sigma(\omega^U)$ is regarded as a homogeneous scalar field in space.

To extend these models to a spatially non-dispersive but anisotropic hysteretic, purely ferromagnetic or ferreolectric  medium it is necessary to accommodate non-local differential constitutive relations with memory.

{\color{black}
In any attempt to model the macroscopic behavior of anisotropic hysteretic material in external time-dependent electromagnetic fields some reliance on experimental data becomes inevitable.
Regretfully phenomenological information on the response of ferromagnetic and ferroelectric materials at rest, to both static and harmonic time-dependent driving fields is sparse.
Reliance on a fair degree of empiricism becomes necessary.
Our approach in large measure is motivated by such sparsity of data.
Consequently a few basic tenets are introduced and these are developed by generalizations that lead to feasible results without excessive numerical computational demands.
These tenets include the requirement that electromagnetic anisotropic responses of the media under consideration can be accommodated in terms of a collection of susceptibility tensors with components that depend nonlinearly on external driving fields.

The dielectric response of many non-hysteretic stationary materials to static external electric fields is coded into their dielectric permittivity tensors.
In general there are three distinct non-coplanar spatial directions in a large specimen of a homogeneous dielectric in which the induced electric polarisation is collinear with the electric field in the medium.
For loss-free dielectrics these directions are mutually orthogonal and constitute frames in which the dielectric permittivity matrix is diagonal.
In any arbitrary stationary orthogonal laboratory frame that is rotated relative to this frame the permittivity matrix will be symmetric.

Similar considerations also apply to stationary materials that possess paramagnetic or diamagnetic susceptibilities.
Stationary ferromagnetic and ferroelectric substances are fundamentally different.
Not only can they acquire a saturated permanent magnetic and electric polarisation respectively but may in general exhibit several (possibly co-planar) preferred directions for their induced magnetic or electric polarization.
Stationary magneto-electric media can also acquire both electric and magnetic polarizations in response to either electric or magnetic driving fields.

A model for a uni-dimensional, rate-independent ferromagnetic medium was constructed by Coleman and Hodgdon \cite{coleman1}, \cite{coleman2}.
In the notation of this article it took the form:
\begin{equation}
\dot{{\cal B}}_{z}=\alpha\left\vert \dot{{\cal H}}_{z}\right\vert \left(
f_1({\cal H}_{z})-{\cal B}_{z}\right) +\dot{{\cal H}}_{z}g_1({\cal H}_{z})
\end{equation}
where the magnetic field component  ${\cal H}_{z}$ and the magnetic induction field component
${\cal B}_{z}$ were continuous slowly varying real-valued functions of time with piecewise continuous
time derivatives, $\dot{{\cal H}}_{z}$, $\dot{{\cal B}} _{z}$, positive constant $\alpha$
 and  $f_1$, $g_1$ specified real-valued functions on the real line.
Following the approach described in the introduction one seeks a covariant extension applicable to ferroelectrics and ferromagnets  that reduces for slowly varying electric or magnetic fields in any inertial frame to a similar type of Coleman-Hodgdon model.
Since such a moving ferromagnet can acquire an induced  electric polarisation and such a moving ferroelectric can acquire a magnetic polarisation, a more general inertial model can be constructed that exhibits a hysteretic  intrinsic magneto-electric response to slowly varying fields in any {\it inertial} frame.
Needless to say there can be no unique extension to accelerating media.

To construct such covariant models we introduce a number of degree $(1,1)$ $V-$orthogonal spatial tensors {\it on Minkowski spacetime}.
These tensors should characterize the response of the material to the fields $F$ and $G$ in an arbitrary spacetime frame $V$.
We restrict to materials that initially possess no permanent electric or magnetic polarisation and have internal microstructure that endows macroscopic media with preferred spatial directions of induced electric and magnetic polarisation.
Such directions are sometimes referred to as `soft' polarisation directions.
Each such tensor ${\cal X}^V$ maps spacetime vector fields to spacetime vector fields, satisfies ${\cal X}^V(V,-)=0$ and ${\cal X}^V(-,\tilde V)=0$ for a time-like vector field $V$.
Furthermore it is supposed that the electric polarisation induced by any electric $1-$form is in the {\it single} direction $N^V_{pe}$ in the frame $V$ while that induced by the magnetic $1-$form is in the {\it single} direction $N^V_{ph}$ in the frame $V$.
Similarly it will assumed that the magnetic polarisation induced by any electric $1-$form is in the {\it single} direction $N^V_{me}$ in the frame $V$ while that induced by the magnetic $1-$form is in the {\it single} direction $N^V_{mh}$ in the frame $V$.
Thus in terms of the unit space-like vector field $N^V_Q$, $(\widetilde{N^V_Q}(N^V_Q)=1)$ with $Q=pe,ph,me,mh$ we write
\begin{equation}
{\cal X}^V_Q={\cal X}_{N^V_Q}\widetilde{{N^V_Q}}\otimes {N^V_Q}  \label{equ:FEtensor}
\end{equation}
with scalar component ${\cal X}_{N^V_Q}$.
}
In terms of ${\cal X}^V_{pe}$, ${\cal X}^V_{ph}$, ${\cal X}^V_{mh}$, ${\cal X}^V_{me}$ an {\it intrinsic magneto-electric} hysteretic constitutive model that has the required inertial behavior  takes the form
\begin{eqnarray}
\nabla _{V}\Pi &=&{\cal X}^V_{pe}\left( \nabla _{V}i_{V}F\right) \wedge
\widetilde{V}+{\cal X}^V_{ph}\left( \nabla _{V}i_{V}\star G\right) \wedge
\widetilde{V}-   \nonumber \\
&&\star \left\{ {\cal X}^V_{mh}\left( \nabla _{V}i_{V}\star G\right) \wedge
\widetilde{V}+{\cal X}^V_{me}\left( \nabla _{V}i_{V}F\right) \wedge
\widetilde{V}\right\}  \label{equ:master1}
\end{eqnarray}
where $\nabla_V$ denotes a covariant derivative with respect to $V$.
If all the tensors ${\cal X}^V_Q$ are non-zero for all $V$ then the material is said to be totally intrinsically magneto-electric.

Since $\nabla _{V}\Pi \neq \nabla _{V}\star \Pi $ when the (Levi-Civita)  $4-$acceleration $\nabla _{V}V$ of the medium is non-zero,
 there exists a physically distinct \lq\lq dual" constitutive model
\begin{eqnarray}
\nabla _{V}\star \Pi &=&{\cal X}^V_{mh}\left( \nabla _{V}i_{V}\star G\right)
\wedge \widetilde{V}+{\cal X}^V_{me}\left( \nabla _{V}i_{V}F\right) \wedge
\widetilde{V}+   \nonumber \\
&&\star \left\{ {\cal X}^V_{pe}\left( \nabla _{V}i_{V}F\right) \wedge
\widetilde{V}+{\cal X}^V_{ph}\left( \nabla _{V}i_{V}\star G\right) \wedge
\widetilde{V}\right\}  \label{equ:master2}
\end{eqnarray}
that has the same behavior as that described by (\ref{equ:master1}) for media at rest in all inertial frames.

The hysteretic behavior of the medium is determined by parameterising the components ${\cal X}_{N_Q^V}$ of each spatial tensor.
%\begin{equation}
%{\cal X} _{Q}^{V}={\cal X} _{N_Q^V}\widetilde{{N}} _{Q}^{V}\otimes {N}_{Q}^{V}  \label{equ:FEtensor}
%\end{equation}
To accommodate such behavior for media in inertial frames the scalar components are written
\begin{eqnarray}
{\cal X} _{N_{pe}^V} &=&\Psi _{N_{pe}^V}\left[ \kappa ^{pe}\mathrm{sgn} \left( \Psi _{N_{pe}^V}\right) +\theta ^{pe}\mathrm{sgn}\left(             \nabla _{V}{\cal E}^{V}({N}_{pe}^{V}) \right) \right] , \\
{\cal X} _{N_{ph}^V} &=&\Psi _{N_{ph}^V}\left[ \kappa ^{ph}\mathrm{sgn} \left( \Psi _{N_{ph}^V}\right) +\theta ^{ph}\mathrm{sgn}\left(  \frac{1}{c}\nabla _{V}{\cal H}^{V}({N}_{ph}^{V}) \right) \right] , \\
{\cal X} _{N_{mh}^V} &=&\Psi _{N_{mh}^V}\left[ \kappa ^{mh}\mathrm{sgn} \left( \Psi _{N_{mh}^V}\right) +\theta ^{mh}\mathrm{sgn}\left(  \frac{1}{c}\nabla _{V}{\cal H}^{V}({N}_{mh}^{V}) \right) \right] , \\
{\cal X} _{N_{me}^V} &=&\Psi _{N_{me}^V}\left[ \kappa ^{me}\mathrm{sgn} \left( \Psi _{N_{me}^V}\right) +\theta ^{me}\mathrm{sgn}\left(             \nabla _{V}{\cal E}^{V}({N}_{me}^{V}) \right) \right]
\end{eqnarray}%
with ${\cal E}^{V}=i_{V}F$, ${\cal H}^{V}=-ci_{V}\star G$ and constants $\kappa ^{Q},\theta ^{Q}\in   \mathbb{R}$.

The $  \mathrm{sgn} $ function  defined by
\begin{equation}
\mathrm{sgn}(z)= z/\vert z \vert = \left\{
\begin{array}{c}
+1\quad for\quad z>0 \\
\quad 0 \quad for \quad z=0 \\
-1\quad for \quad z<0
\end{array}
\right.
\end{equation}
accommodates the {\it branching} during the hysteretic process with
\begin{eqnarray}
\Psi _{N_{pe}^V} &=&\epsilon _{0}\left( \alpha ^{pe}f^{pe}({\cal E}_{N_{pe}^V})-\xi ^{pe}{\cal P}_{N_{pe}^V}\right) \label{equ:PsiNpeV} \\
\Psi _{N_{ph}^V} &=&-            \left( \alpha ^{ph}f^{ph}({\cal H}_{N_{ph}^V})-\xi ^{ph}{\cal P}_{N_{ph}^V}\right) \\
\Psi _{N_{mh}^V} &=&-            \left( \alpha ^{mh}f^{mh}({\cal H}_{N_{mh}^V})+\xi ^{mh}{\cal M}_{N_{mh}^V}\right) \\
\Psi _{N_{me}^V} &=&\epsilon _{0}\left( \alpha ^{me}f^{me}({\cal E}_{N_{me}^V})+\xi ^{me}{\cal M}_{N_{me}^V}\right),
\label{equ:FMcom}
\end{eqnarray}
constants $\alpha ^{Q},\xi ^{Q}\in   \mathbb{R}  ^{+}$ and frame-dependent scalars
\begin{eqnarray}
{\cal E}_{N_Q^V}&=&  i_{V}F  (N_Q^V),  \quad \quad \quad  {\cal H}_{N_Q^V}=-ci_{V}\star G  (N_Q^V),\\
{\cal P}_{N_Q^V}&=&  i_{V}\Pi(N_Q^V),  \quad \quad \quad  {\cal M}_{N_Q^V}=-ci_{V}\star \Pi(N_Q^V)
\end{eqnarray}%
which are the components of ${\cal E}^V,{\cal H}^V,{\cal P}^V,{\cal M}^V$ projected onto the directions $N_Q^V$ respectively.
(Note, the factors of $\epsilon_0$ imply that $\Psi _{N_{pe}^V}/\epsilon _{0},\Psi _{N_{ph}^V},\Psi _{N_{me}^V}/\epsilon _{0},\Psi _{N_{mh}^V}$
are dimensionless.)

Each real valued  function $f^Q$ characterizes the {\it details} of the hysteretic process and is chosen to be:
\begin{equation}
f^Q(z)=\tanh ( \beta^Q z)
\label{equ:sigmoid}
\end{equation}
with $\beta^Q \in  \mathbb{R}  ^{+}$ so that $ \lim_{z\rightarrow \pm \infty}
f^Q (z)=\pm 1$. The values of the parameters $\alpha^Q, \beta^Q,\xi^Q\,\kappa^Q,\theta^Q$ must be motivated by available data in the laboratory frame.

The use of the  covariant derivative in (\ref{equ:master1}) and (\ref{equ:master2}) ensures that a hysteretic process
driven by a harmonic external field can  give rise to an autonomous differential equation on any
{\it smooth branch} of a hysteresis loop as described in the next section.
% (see (\ref{equ:hysPE}) and (\ref{equ:hysMH})).
In any inertial frame $U$ both  (\ref{equ:master1}) and (\ref{equ:master2}) yield the coupled
system of differential equations:
\begin{eqnarray}
\dot{\cal P}^U&=&{\cal X} _{pe}^U\left( \dot{\cal E}^U\right) -\frac{1}{c} {\cal X} _{ph}^U \left( \dot{\cal H}^U\right)  \\
\dot{\cal M}^U&=&{\cal X} _{mh}^U\left( \dot{\cal H}^U\right) -c{           \cal X} _{me}^U \left( \dot{\cal E}^U\right)
\end{eqnarray}
where
$\dot{\cal E}^U \equiv \nabla _U {\cal E}^U$,
$\dot{\cal H}^U \equiv \nabla _U {\cal H}^U$,
$\dot{\cal P}^U \equiv \nabla _U {\cal P}^U$ and
$\dot{\cal M}^U \equiv \nabla _U {\cal M}^U$.

\section{A Particular Model at Rest in an Inertial (Laboratory)  Frame}

We  consider (\ref{equ:master1}) for the special case of a particular non-magneto-electric medium where ${\cal X}_{ph}^V=0$ and ${\cal X} _{me}^V=0$. Then
\begin{equation}
\nabla _{V}\Pi ={\cal X} _{pe}^V\left( \nabla _{V}i_{V}F\right) \wedge \widetilde{V}-\star \left( {\cal X} _{mh}^V\left( \nabla _{V}i_{V}\star G\right) \wedge \widetilde{V}\right)  \label{equ:covar1}
\end{equation}
 In terms of spatial fields it follows from (\ref{equ:PI}) that (\ref {equ:covar1}) may be written,
\begin{eqnarray}
\nabla _{V}{\cal P}^{V}\wedge \widetilde{V}&+&{\cal P}^{V}\wedge \nabla _{V}\widetilde{V} -\nabla _{V}\left[\star \left( \frac{{\cal M}^{V}}{c}\wedge \widetilde{V}\right)\right] =   \nonumber \\
&& {\cal X}_{pe}^V \left( \nabla _{V}{\cal E}^{V}\right) \wedge \widetilde{V}+\star
\left( {\cal X} _{mh}^V \left( \nabla _{V}\frac{{\cal H}^{V}}{c}\right) \wedge
\widetilde{V}\right)  \label{equ:covar2}
\end{eqnarray}
For the medium at rest in the inertial frame  $U=\frac{1}{c}\partial _{t}$ one has $V=U$
and  equation (\ref{equ:covar2}) becomes
\begin{equation}
\dot{{\cal P}}^{U}\wedge \widetilde{U}+\star \left( \frac{\dot{{\cal M}}^{U}}{c}\wedge
\widetilde{U}\right) = {\cal X} _{pe}^U\left( \dot{{\cal E}}^{U}\right) \wedge
\widetilde{U}+\star \left( {\cal X} _{mh}^U\left( \frac{\dot{{\cal H}}^{U}}{c}\right)
\wedge \widetilde{U}\right)
\end{equation}
where for any spatial $p-$form $\xi ^{U}$
\begin{equation}
\dot{\xi}^{U} = \nabla _{\partial _{t}}\xi ^{U}
\end{equation}
since the (laboratory) inertial cobasis is parallel.
Taking components in the direction of $U$ and its orthogonal subspace yields
\begin{eqnarray}
\dot{{\cal P}}^{U} &=&{\cal X} _{pe}^U\left( \dot{{\cal E}}^{U}\right) \\
\dot{{\cal M}}^{U} &=&{\cal X} _{mh}^U\left( \dot{{\cal H}}^{U}\right)
\end{eqnarray}
One now has in the $U$ frame  decoupled equations for the hysteretic electric and magnetic polarisation fields in the medium:
\begin{eqnarray}
\dot{{\cal P}}_{N_{pe}^U} &=&\kappa ^{pe}\dot{{\cal E}}_{N_{pe}^U}\left\vert \Psi _{N_{pe}^U}\right\vert +\theta ^{pe}\left\vert \dot{{\cal E}}_{N_{pe}^U}\right\vert \Psi _{N_{pe}^U}  \label{equ:Updot} \\
\dot{{\cal M}}_{N_{mh}^U} &=&\kappa ^{mh}\dot{{\cal H}}_{N_{mh}^U}\left\vert \Psi _{N_{mh}^U}\right\vert +\theta ^{mh}\left\vert \dot{{\cal H}}_{N_{mh}^U}\right\vert \Psi _{N_{mh}^U}  \label{equ:Umdot}
\end{eqnarray}
where
$\dot{\cal P}_{N_{pe}^U}=\dot{\cal P}^U(N_{pe}^U)$,
$\dot{\cal E}_{N_{pe}^U}=\dot{\cal E}^U(N_{pe}^U)$,
$\dot{\cal M}_{N_{mh}^U}=\dot{\cal M}^U(N_{mh}^U)$ and
$\dot{\cal H}_{N_{mh}^U}=\dot{\cal H}^U(N_{mh}^U)$.

Over an {\it arbitrary} time interval each of the above  differential equations is a non-autonomous evolution equation at each point in the medium  describing  induced polarisations as a function of time.
 However in certain time domains their evolution is controlled    by autonomous ordinary differential equations.
 When the drive fields are harmonic in time the resulting solutions to such equations  may exhibit a limit cycle, (often identified as a hysteresis loop) composed of piecewise smooth line segments.
 For such cycles  controlled by (\ref{equ:Updot}) the location and number of piecewise smooth segments in the cycle (branches) are determined by the location of the zeroes of $\dot{{\cal E}}_{N_{pe}^U}$ and  $\Psi_{N_{pe}^U}$ in the cycle.
 Similarly in cycles  controlled by (\ref{equ:Umdot}) branches are determined by the location of the zeroes of $\dot{{\cal H}}_{N_{mh}^U}$ and  $\Psi_{N_{mh}^U}$ in the cycle.
 The choice of parameters in the functions $\Psi_{N_{pe}^U}$ and $\Psi_{N_{mh}^U}$ determine such locations and also the degree of induced saturation in each branch of a limit cycle during the process.
  When $ \dot{{\cal E}}_{N_{pe}^U}$  and  $ \dot{{\cal H}}_{N_{mh}^U}$  have more general time-dependences the above equations give rise to solutions that may exhibit hysteretic loci containing self-intersections and/or no limit cycle.

Thus at each spatial point, (\ref{equ:Updot}) can be written

\begin{equation}
\frac{\dot{{\cal P}}_{N_{pe}^U}}{\dot{{\cal E}}_{N_{pe}^U}}=\frac{d{\cal P}_{N_{pe}^U}}{d{\cal E}_{N_{pe}^U}}
=\kappa ^{pe} \Psi _{N_{pe}^U} \mathrm{sgn}(\Psi _{N_{pe}^U}) +\theta
^{pe}\mathrm{sgn}(\dot{{\cal E}}_{N_{pe}^U})\Psi _{N_{pe}^U}  \label{equ:hysPE}
\end{equation}
For each of the possible combinations of $\mathrm{sgn}(\dot{{\cal E}}_{N_{pe}^U})$ and $\mathrm{sgn}(\Psi _{N_{pe}^U})$, (\ref%
{equ:hysPE}) yields a first-order differential
equation for ${\cal P}_{N_{pe}^U}({\cal E}_{N_{pe}^U})$ with branched solutions through any point $({\cal P}_{0}^{U},{\cal E}_{0}^{U})$.
If one denotes
$\mathrm{sgn}(\dot{{\cal E}}_{N_{pe}^U})$ by $S_E$ and $\mathrm{sgn}(\Psi_{N_{pe}^U})$ by $S_\Psi$
and furthermore assumes that parameters are chosen so that $S_\Psi$ is constant on a particular branch a solution through this point can be written
\begin{eqnarray}
{\cal P}_{N_{pe}^U}({\cal E}_{N_{pe}^U};{\cal P}_{0}^{U},{\cal E}_{0}^{U},S_{\Psi },S_{E}) ={\cal P}_{0}^{U}e^{-(\eta ^{pe}\xi
^{pe}({\cal E}_{N_{pe}^U}-{\cal E}_{0}^{U}))}+
\nonumber \\
\quad\quad\quad\quad\quad e^{-(\eta^{pe}\xi ^{pe}{\cal E}_{N_{pe}^U})}\eta^{pe}\alpha
^{pe}\int_{{\cal E}_{0}^{U}}^{{\cal E}_{N_{pe}^U}}f^{pe}\left( \Upsilon \right) e^{(\eta^{pe}\xi
^{pe}\Upsilon )}d\Upsilon
\end{eqnarray}
where $\eta ^{pe}=\epsilon _{0}\left( \kappa ^{pe}S_{\Psi }+\theta ^{pe}S_{E}\right) $.
Although $S_E$ is $+1$ when $\dot {\cal E}_{N_{pe}^U}$ is increasing with time or $-1$ when $\dot {\cal E}_{N_{pe}^U}$ is
decreasing with time the sign of $S_\Psi$ will in general depend on the state $({\cal P}_{N_{pe}^U},{\cal E}_{N_{pe}^U})$. Furthermore when the parameters $\alpha^{pe}, \beta^{pe},\xi^{pe}, \kappa^{pe}$  and $\theta^{pe}$ vary inhomogeneously with position in the medium this state will also depend explicitly on position. In such
situations an analytic solution is no longer in general possible. Analogous solutions may be written for
(\ref{equ:Umdot}).

\section{A Particular Model Rotating in an Inertial (Laboratory)  Frame}

\begin{figure}[!t]
\centering
\includegraphics[scale=1]{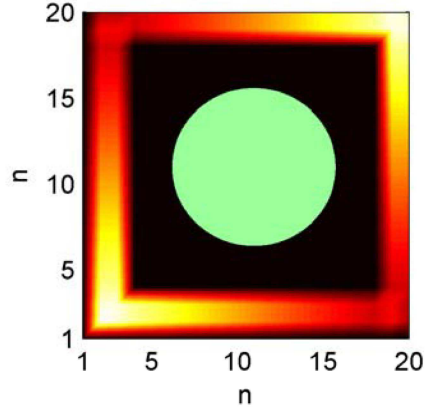}
\caption{The computational domain is the space inside a perfectly conducting rectangular cavity with dimensions $5.45$m$
\times 5.45$m$ \times 2.18$m.  This figure shows an $x-y$  cross-section of the cavity at constant $z$. The computation is done on a $20 \times 20 \times 8$ 3-dimensional lattice in the cavity. The fields in the cavity are driven by a spatially divergence-free time-harmonic current density  that flows parallel to the pairs of $x-y$, $y-z$ and $z-x$   interior faces of the cavity without penetrating the cylindrical ferroelectric material with radius $1.36$m, centered at the origin, oriented with its axis of rotation along the $z-$axis of the cavity.
  % A time-harmonic current density is taken proportional to the spatial vector with Cartesian components $(z,z,x)$ modulated to have support that excludes the cylinder. In the figure the gradation of shading from dark to % light around the  $x-y$  perimeter depicts,  from minimum to maximum, the magnitude of the current density in the cross section  at a particular instant of time.
  }
\label{figure1}
\end{figure}

In this section the model is applied to a rigid uniformly rotating cylinder of radius $R$
in an external time harmonic electromagnetic field in an effort to see the magnitude of effects for a
hysteretic process in an accelerating medium. The differential constitutive relations are appended
to the Maxwell system and the resulting coupled differential system analyzed numerically.
The system is wholly enclosed in a rectangular 3-dimensional  computational domain that
simulates the rotating cylinder in a perfectly conducting vacuum cavity; see figure \ref{figure1}. The choice of parameters in the model is
motivated by establishing results that may be compared with hysteretic behavior of a medium at rest in a low frequency electromagnetic environment.
 The modern  measurement of ferromagnetic and ferroelectric hysteresis  in such situations  is an experimental art and owes much to modern digital and piezoelectric technology. Needless to say  great care is exercised to accommodate effects such as sample geometry,  material electrical conductivity variations with frequency, thermal drift instabilities and material spatial inhomogeneities \cite{data}. In this article we will use a particular hysteresis measurement \cite{boch}  more as a guide to a reasonable parameter set for our model rather than a targeted fit to a specific material specimen.

The  model is applied to a  $1.36$m  radius
cylinder, rotating at an angular speed  $1497$rpm inside a  perfectly conducting cavity with sides of lengths $5.45$m$\times 5.45$m$ \times 2.18$m.  The cavity fields are  driven by an  assembly of
externally prescribed  currents  inside the cavity at  $250$Hz, well below the lowest
natural electromagnetic mode of such a cavity.
The associated current density is restricted to a few mesh intervals in the vicinity of the interior cavity wall and is prescribed to be divergence-free for all time.
For the above angular speed and cavity dimension
a lattice $20 \times 20 \times 8   $  discretization  (using a fast workstation)  permits exploration of the hysteretic
processes in the rotating medium when the above differential equations are coupled to the macroscopic Maxwell equations. In this approach all components of spatial electromagnetic fields and spatial polarisations are
with respect to the laboratory frame $U$ and interest is directed to how these fields evolve parametrically with time and hence with each other in such a rotating medium.

We consider a purely ferroelectric medium that is  magnetically inert with no initial permanent polarization in its rest frame subject to the constitutive relation
\begin{eqnarray}
\nabla _{V}\Pi &=&{\cal X}^V_{pe}\left( \nabla _{V}i_{V}F\right) \wedge
\widetilde{V} \label{equ:pemodel}
\end{eqnarray}
with
\begin{eqnarray}
{\cal X}^V_{pe}={\cal X}_{N^V_{pe}} {\widetilde {N^V_{pe}}} \otimes N^V_{pe}
\end{eqnarray}
and ${\cal X}_{N^V_{pe}}$ parameterized as in (\ref{equ:PsiNpeV}).

The cylinder will be set in uniform rotary motion with angular speed $\Omega$ radians per second about the $z$-axis of a laboratory Cartesian frame.
At this point it is natural to introduce a cylindrical polar Minkowski cobasis with $e^{0}=cdt,$ $e^{1}=dr$, $e^{2}=rd\phi$, $e^{3}=dz$ and
\begin{equation}
\mathbf{g}=-c^{2}dt\otimes dt+r^{2}d\phi \otimes d\phi +dr\otimes dr+dz\otimes dz
\end{equation}
where
\begin{eqnarray}
r^{2} &=&x^{2}+y^{2},\quad x=r\cos \phi ,\quad y=r\sin
\phi  \\
\partial _{\phi } &=&x\partial _{y}-y\partial _{x},\quad \partial _{r}=\frac{1}{r}\left( x\partial _{x}+y\partial _{y}\right)
 \\
d\phi  &=&\frac{1}{r^{2}}\left( xdy-ydx\right) ,\quad dr=\frac{1 }{r}\left(
xdx+ydy\right)
\end{eqnarray}
As a vector field on spacetime, the bulk 4-velocity field of the rotating cylinder in Minkowski cylindrical polar coordinates is
\begin{equation}
V=\frac{\gamma }{c}\left( \partial _{t}+\Omega \partial _{\phi }\right) \quad with \quad \gamma =\left( 1-\frac{r^{2}\Omega ^{2}}{ c^{2}}\right) ^{-\frac{1}{2}}
\end{equation}
Since $r\Omega \ll c$ for $r\leqslant R$,
\begin{eqnarray}
V &\simeq &\frac{1}{c}\partial _{t}+\frac{\Omega }{c}\left( x\partial
_{y}-y\partial _{x}\right)  \\
\widetilde{V} &\simeq &-cdt+\frac{\Omega }{c}\left( xdy-ydx\right)
\end{eqnarray}
and $\mathbf{g}(V,V)\simeq r^{2}\frac{\Omega ^{2}}{c^{2}}-1\approx -1$.
The `soft' direction of polarisation can be aligned with the vector field $\partial_z$ since $\mathbf{g}(V,\partial_z)=0$ so that
\begin{eqnarray}
{\cal X}^V_{pe}={\cal X}_{\partial_z} \, dz \otimes \partial_z
\end{eqnarray}

\def\cds{ (t,x,y,z) }

Using the relations
\begin{eqnarray}
F &=& {\cal E}^{U} \wedge \widetilde{U} + c B^U = {\cal E}^{V} \wedge \widetilde{V} + c B^V,\\
{\cal E}^{V} &=& i_V F,\\
{\cal E}^{U} &=& i_U F,\\
V & \approx & U+\frac{\Omega }{c}\left( x \partial_y-y \partial_x\right),
\end{eqnarray}
all fields can  now be  projected into the laboratory $U$ frame,  terms of order $\left( r\Omega \right) ^{2}/c^{2}$
removed and components expressed relative to the Minkowski cobasis in $t,x,y,z$ coordinates where
\begin{equation}
U =\frac{1}{c} \partial _{t}, \quad
\widetilde{U}=-c dt
\end{equation}
and $\# 1= dx \wedge dy \wedge dz$.

Thus reverting to laboratory Cartesian coordinates and noting that $\nabla _{V}\widetilde{U}=0$, $\nabla _{U}\widetilde{U}=0$, the left hand side (\ref{equ:pemodel}) is
\begin{eqnarray}
\nabla _{V}\Pi &=&\nabla _{U+\frac{\Omega }{c}\left( x\partial _{y}-y\partial _{x}\right) }\left(
{\cal P}^{U}\wedge \widetilde{U}-\frac{M^{U}}{c}
\right)   \nonumber \\
&=&\left( U{\cal P}^{U}+\frac{\Omega }{c}\left( x\partial _{y}-y\partial
_{x}\right) {\cal P}^{U}\right) \wedge \widetilde{U}-\frac{1}{c}UM^{U}- \nonumber \\
&& \frac{\Omega}{c^{2}}\left( x\partial _{y}-y\partial _{x}\right) M^{U}
\end{eqnarray}
(with $\Pi={\cal P}^{U} \wedge \widetilde{U} - \frac{1}{c} M^U$,  $M^{U}=\# {\cal M}^{U}$)
and the right hand side of (\ref{equ:pemodel}) is
\begin{eqnarray}
\ {\cal X} _{pe}^V\left( \nabla _{V}i_{V}F\right) \wedge \widetilde{V} =
{\cal X} _{pe}^V\left( \nabla _{V}{\cal E}^{V}\right) \wedge \left( \widetilde{U}+
\frac{\Omega }{c}\left( xdy-ydx\right) \right).
\end{eqnarray}

The six components of (\ref{equ:pemodel}) yield the
following coupled system for the functions  $ {{\cal P}}_{x}^{U} \cds$, $ {{\cal P}}_{y}^{U} \cds$, ${{\cal P}}_{z}^{U} \cds$, ${{\cal M}}_{x}^{U} \cds$, ${{\cal M}}_{y}^{U} \cds$, ${{\cal M}}_{z}^{U} \cds$
where ${\cal P}^{U}_x={\cal P}^{U}(\partial_{x})$, etc, ${\cal M}^{U}_x={\cal M}^{U}(\partial_{x})$, etc:

\bigskip
\textit{Electric Sector}
\begin{eqnarray}
\dot{\mathcal{P}}_{x}^{U}+\Omega \left( x\partial _{y}-y\partial _{x}\right)
\mathcal{P}_{x}^{U} &=&0  \label{equ:sysbgn} \\
\dot{\mathcal{P}}_{y}^{U}+\Omega \left( x\partial _{y}-y\partial _{x}\right)
\mathcal{P}_{y}^{U} &=&0 \\
\dot{\mathcal{P}}_{z}^{U}+\Omega \left( x\partial _{y}-y\partial _{x}\right)
\mathcal{P}_{z}^{U} &=&\mathcal{X}_{\partial_z}\left\{ \dot{\mathcal{E}}%
_{z}^{U}-\Omega \left[ x\dot{\mathcal{B}}_{x}^{U}+y\dot{\mathcal{B}}_{y}^{U}%
\right] +\Omega \left( x\partial _{y}-y\partial _{x}\right) \mathcal{E}%
_{z}^{U}\right\}
\end{eqnarray}

\textit{Magnetic Sector}
\begin{eqnarray}
\dot{{\cal M}}_{x}^{U}+\Omega \left( x\partial _{y}-y\partial _{x}\right) {\cal M}_{x}^{U}
&=&x\Omega \, {\cal X} _{\partial_z}\dot{{\cal E}}_{z}^{U} \label{equ:mxxx}\\
\dot{{\cal M}}_{y}^{U}+\Omega \left( x\partial _{y}-y\partial _{x}\right) {\cal M}_{y}^{U}
&=&y\Omega \, {\cal X} _{\partial_z}\dot{{\cal E}}_{z}^{U} \label{equ:myyy}\\
\dot{{\cal M}}_{z}^{U}+\Omega \left( x\partial _{y}-y\partial _{x}\right) {\cal M}_{z}^{U} &=&0 \label{equ:mzzz}
\end{eqnarray}%
with
\begin{eqnarray}
{\cal X} _{\partial_z} &=&\epsilon _{0}\left\{ \alpha^{pe}f^{pe}\left(
{\cal E}_{z}^{U}-\Omega \left[ x{\cal B}_{x}^{U}+y{\cal B}_{y}^{U}\right] \right) -\xi^{pe}
\left[ {\cal P}_{z}^{U}-\Omega \frac{x{\cal M}_{x}^{U}+y{\cal M}_{y}^{U}}{{c}^{2}}\right]
\right\}.   \nonumber \\
&&\left\{ \kappa ^{pe}\mathrm{sgn}\left( \epsilon _{0}\left\{ \alpha
^{pe}f^{pe}\left( {\cal E}_{z}^{U}-\Omega \left[
x{\cal B}_{x}^{U}+y{\cal B}_{y}^{U}\right] \right) -\xi ^{pe}\left[ {\cal P}_{z}^{U}{-}\Omega
\frac{x{\cal M}_{x}^{U}+y{\cal M}_{y}^{U}}{{c}^{2}}\right] \right\} \right) +\right.
\nonumber \\
&&\left. \theta ^{pe}\mathrm{sgn}\left( \dot{{\cal E}}_{z}^{U}-\Omega \left[
y\,\dot{{\cal B}}_{y}^{U}+x\,\dot{{\cal B}}_{x}^{U}\right] +\Omega \left( \,x{\partial }%
_{y}-\,y{\partial }_{x}\right) {\cal E}_{z}^{U}\right) \right\} \label{equ:sysend}
\end{eqnarray}%
As discussed in Section 4 the presence of the discontinuous $\mathrm{sgn}$ functions in these equations is responsible for the branched structure of their solutions.
Substituting (\ref{equ:sysend}) into (\ref{equ:sysbgn})-(\ref{equ:mzzz}) then yields the coupled system
\begin{eqnarray}
\dot{{\cal P}}_{x}^{U} &=&\Omega \left( y\partial _{x}-x\partial _{y}\right)
{\cal P}_{x}^{U}  \label{equ:Omgpx} \\
\dot{{\cal P}}_{y}^{U} &=&\Omega \left( y\partial _{x}-x\partial _{y}\right)
{\cal P}_{y}^{U}  \label{equ:Omgpy} \\
\dot{{\cal P}}_{z}^{U} &=&\Omega \left( y\partial _{x}-x\partial _{y}\right)
{\cal P}_{z}^{U}+\epsilon _{0}\left\{ \alpha ^{pe}f^{pe}(\hat e_{3}^{V})-\xi
^{pe}{\hat {\cal P}}_{3}^{V}\right\} .  \nonumber \\
&&\left\{ \kappa ^{pe}\hat {\mathcal{E}}_{3}^{V}\mathrm{sgn}\left( \epsilon _{0}%
\left[ \alpha ^{pe}f^{pe}\left( \hat e_{3}^{V}\right) -\xi
^{pe} \hat{\cal P}_{3}^{V}\right] \right) +\theta ^{pe}\left\vert \hat{ \mathcal{E}}%
_{3}^{V}\right\vert \right\}   \label{equ:Pzdot} \\
\dot{{\cal M}}_{x}^{U} &=&\Omega \left( y\partial _{x}-x\partial _{y}\right)
{\cal M}_{x}^{U}+   \nonumber \\
&&+x\Omega \epsilon _{0}\kappa ^{pe}\left\vert \alpha
^{pe}f^{pe}({\cal E}_{z}^{U})-\xi ^{pe}{\cal P}_{z}^{U} \right\vert
\dot{{\cal E}}_{z}^{U}+  \nonumber \\
&&+x\Omega \epsilon _{0}\theta ^{pe}\left[ \alpha
^{pe}f^{pe}({\cal E}_{z}^{U})-\xi ^{pe}{\cal P}_{z}^{U}\right] \left\vert \dot{{\cal E}}_{z}^{U}\right\vert \label{equ:Mxdot} \\
\dot{{\cal M}}_{y}^{U} &=&\Omega \left( y\partial _{x}-x\partial _{y}\right)
{\cal M}_{y}^{U}+  \nonumber \\
&&+ y\Omega \epsilon _{0}\kappa ^{pe}\left\vert \alpha
^{pe}f^{pe}({\cal E}_{z}^{U})-\xi ^{pe}{\cal P}_{z}^{U} \right\vert
\dot{{\cal E}}_{z}^{U}+ \nonumber \\
&&+y\Omega \epsilon _{0}\theta ^{pe}\left[ \alpha
^{pe}f^{pe}({\cal E}_{z}^{U})-\xi ^{pe}{\cal P}_{z}^{U}\right] \left\vert \dot{{\cal E}}_{z}^{U}\right\vert  \label{equ:Mydot} \\
\dot{{\cal M}}_{z}^{U} &=&\Omega \left( y\partial _{x}-x\partial _{y}\right)
{\cal M}_{z}^{U}  \label{equ:Omgmz}
\end{eqnarray}%
where we define
\begin{eqnarray}
\hat {\cal P}_{3}^{V} &=&{\cal P}_{z}^{U}-\frac{\Omega }{c^{2}}[x{\cal M}_{x}^{U}+y{\cal M}_{y}^{U}]
\label{equ:Pv3} \\
\hat e_{3}^{V} &=&{\cal E}_{z}^{U}-\Omega [x{\cal B}_{x}^{U}+y{\cal B}_{y}^{U}]  \label{equ:Ev3} \\
\hat{\mathcal{E}}_{3}^{V} &=&\dot{{\cal E}}_{z}^{U}-\Omega \left[ y\,\dot{{\cal B}}_{y}^{U}+x\,%
\dot{{\cal B}}_{x}^{U}\right] +\Omega \left( \,x{\partial }_{y}-\,y{\partial }%
_{x}\right) {\cal E}_{z}^{U}  \label{equ:EEEv3}
\end{eqnarray}%
and terms of order $(r\Omega /c)^{2}$ have been consistently dropped.
Thus when substituting (\ref{equ:sysend}) into (\ref{equ:mxxx}) and (\ref{equ:myyy})
it is sufficient to replace
$\hat{\cal P}_{3}^{V}$ with ${\cal P}_{z}^{U}, \hat e_{3}^{V}$  with $ {\cal E}_{z}^{U}$ and $\hat {\mathcal{E}}_{3}^{V}$ with $\dot{{\cal E}}_{z}^{U}$
yielding (\ref{equ:Mxdot}) and (\ref{equ:Mydot}).
The time derivative of the magnetic field in (\ref{equ:EEEv3}) is found using the time derivative of (\ref{equ:Hconstitutive}).
%From (\ref{equ:Pv3}) this implies that  it is
%sufficient to replace ${\cal P}_{3}^{V}$ with ${\cal P}_{z}^{U}$ in (\ref{equ:Mxdot}) and (\ref{equ:Mydot}). \
%Similarly  (\ref{equ:Ev3}) implies that it is sufficient to replace ${\cal E}_{3}^{V}$ with ${\cal E}_{z}^{U}$ in
%(\ref{equ:Pzdot}), (\ref{equ:Mxdot}) and (\ref{equ:Mydot})
%and,  to order $(r\Omega /c)^{2}$, $\Omega {\cal P}_{3}^{V}\equiv \Omega {\cal P}_{z}^{U}$
%and $\Omega {\cal E}_{3}^{V}\equiv \Omega {\cal E}_{z}^{U}$.

\section{Numerical Analysis}

The above constitutive equations for components of the induced polarisation in the rotating ferroelectric are coupled to the macroscopic Maxwell equations described in section 2.
Since we assume that the ferroelectric medium has no initial polarization, equations (\ref{equ:Omgpx}), (\ref{equ:Omgpy}) and (\ref{equ:Omgmz}) imply that
${\cal P}_x^U=0$, ${\cal P}_y^U=0$ and ${\cal M}_z^U=0$ for all time.
Thus the system when coupled with Maxwell's equations inside the cylinder becomes:
\begin{eqnarray}
\dot{{\cal P}}_{z}^{U} &=&\Omega \left( y\partial _{x}-x\partial _{y}\right)
{\cal P}_{z}^{U}+\epsilon _{0}\left\{ \alpha ^{pe}\tanh(\beta ^{pe} \hat e_{3}^{V})-\xi
^{pe} \hat{\cal P}_{3}^{V}\right\} .  \nonumber \\
&&\left\{ \kappa ^{pe}\hat{\mathcal{E}}_{3}^{V}\mathrm{sgn}\left( \epsilon _{0}%
\left[ \alpha ^{pe}\tanh\left(\beta ^{pe} \hat e_{3}^{V}\right) -\xi
^{pe} \hat{\cal P}_{3}^{V}\right] \right) +\theta ^{pe}\left\vert \hat{\mathcal{E}}%
_{3}^{V}\right\vert \right\}  \label{equ:Pdscld}\\
\dot{{\cal M}}_{x}^{U} &=&\Omega \left( y\partial _{x}-x\partial _{y}\right)
{\cal M}_{x}^{U}+  \nonumber \\
&&x\Omega \epsilon _{0}\kappa ^{pe}\left\vert \alpha
^{pe}\tanh(\beta ^{pe} {\cal E}_{z}^{U})-\xi ^{pe}{\cal P}_{z}^{U} \right\vert
\dot{{\cal E}}_{z}^{U}+  \nonumber \\
&&x\Omega \epsilon _{0}\theta ^{pe}\left[ \alpha
^{pe}\tanh(\beta ^{pe} {\cal E}_{z}^{U})-\xi ^{pe}{\cal P}_{z}^{U}\right] \left\vert \dot{{\cal E}}_{z}^{U}\right\vert   \label{equ:M1scld} \\
\dot{{\cal M}}_{y}^{U} &=&\Omega \left( y\partial _{x}-x\partial _{y}\right)
{\cal M}_{y}^{U}+  \nonumber \\
&&y\Omega \epsilon _{0}\kappa ^{pe}\left\vert \alpha
^{pe}\tanh(\beta ^{pe} {\cal E}_{z}^{U})-\xi ^{pe}{\cal P}_{z}^{U} \right\vert
\dot{{\cal E}}_{z}^{U}+  \nonumber \\
&&y\Omega \epsilon _{0}\theta ^{pe}\left[ \alpha
^{pe}\tanh(\beta ^{pe}{\cal E}_{z}^{U})-\xi ^{pe}{\cal P}_{z}^{U}\right] \left\vert \dot{{\cal E}}_{z}^{U}\right\vert   \label{equ:M2scld}
\end{eqnarray}%
and
\begin{eqnarray}
\epsilon _{0}\dot{{\cal E}}_{x}^{U} &=&(\nabla \times \mathbf{H}^{U})_{x}-\sigma {\cal E}_{x}^{U}  \label{equ:Pscld1MW} \\
\epsilon _{0}\dot{{\cal E}}_{y}^{U} &=&(\nabla \times \mathbf{H}^{U})_{y}-\sigma {\cal E}_{y}^{U} \\
\epsilon _{0}\dot{{\cal E}}_{z}^{U} &=&(\nabla \times \mathbf{H}^{U})_{z}-\sigma {\cal E}_{z}^{U}-\dot{{\cal P}}_{z}^{U}  \label{equ:Pscld2MW} \\
\dot{{\cal H}}_{x}^{U} &=&\frac{-1}{\mu _{0}}(\nabla \times \mathbf{E}^{U})_{x}+ \dot{{\cal M}}_{x}^{U}  \label{equ:M1scldMW} \\
\dot{{\cal H}}_{y}^{U} &=&\frac{-1}{\mu _{0}}(\nabla \times \mathbf{E}^{U})_{y}+ \dot{{\cal M}}_{y}^{U}  \label{equ:M2scldMW} \\
\dot{{\cal H}}_{z}^{U} &=&\frac{-1}{\mu _{0}}(\nabla \times \mathbf{E}^{U})_{z} \label{equ:Hzscld}
\end{eqnarray}

To complete the analysis these equations must be coupled to the vacuum Maxwell system with a current source in the vacuum region between the cavity walls and the surface of the cylinder:

\begin{eqnarray}
\epsilon _{0}\dot{{\cal E}}_{x}^{U} &=&(\nabla \times \mathbf{H}^{U})_{x}-j_{ext,x}^{U}  \label{equ:Pscld1MWA} \\
\epsilon _{0}\dot{{\cal E}}_{y}^{U} &=&(\nabla \times \mathbf{H}^{U})_{y}-j_{ext,y}^{U} \\
\epsilon _{0}\dot{{\cal E}}_{z}^{U} &=&(\nabla \times \mathbf{H}^{U})_{z}-j_{ext,z}^{U}  \label{equ:Pscld2MWA} \\
\dot{{\cal H}}_{x}^{U} &=&\frac{-1}{\mu _{0}}(\nabla \times \mathbf{E}^{U})_{x}   \label{equ:M1scldMWA} \\
\dot{{\cal H}}_{y}^{U} &=&\frac{-1}{\mu _{0}}(\nabla \times \mathbf{E}^{U})_{y}   \label{equ:M2scldMWA} \\
\dot{{\cal H}}_{z}^{U} &=&\frac{-1}{\mu _{0}}(\nabla \times \mathbf{E}^{U})_{z} \label{equ:HzscldA}
\end{eqnarray}
%\begin{eqnarray}
%\epsilon _{0}\dot{{\cal E}}_{x}^{U} &=&(\nabla \times \mathbf{H}^{U})_{x}-\sigma {\cal E}_{x}^{U}-j_{ext,x}^{U}  \label{equ:Pscld1MW} \\
%\epsilon _{0}\dot{{\cal E}}_{y}^{U} &=&(\nabla \times \mathbf{H}^{U})_{y}-\sigma {\cal E}_{y}^{U}-j_{ext,y}^{U} \\
%\epsilon _{0}\dot{{\cal E}}_{z}^{U} &=&(\nabla \times \mathbf{H}^{U})_{z}-\sigma {\cal E}_{z}^{U}-j_{ext,z}^{U}-\dot{{\cal P}}_{z}^{U}  \label{equ:Pscld2MW} \\
%\dot{{\cal H}}_{x}^{U} &=&\frac{-1}{\mu _{0}}(\nabla \times \mathbf{E}^{U})_{x}+ \dot{{\cal M}}_{x}^{U}  \label{equ:M1scldMW} \\
%\dot{{\cal H}}_{y}^{U} &=&\frac{-1}{\mu _{0}}(\nabla \times \mathbf{E}^{U})_{y}+ \dot{{\cal M}}_{y}^{U}  \label{equ:M2scldMW} \\
%\dot{{\cal H}}_{z}^{U} &=&\frac{-1}{\mu _{0}}(\nabla \times \mathbf{E}^{U})_{z} \label{equ:Hzscld}
%\end{eqnarray}

\begin{figure}[!h]
\centering
\includegraphics[scale=2.2]{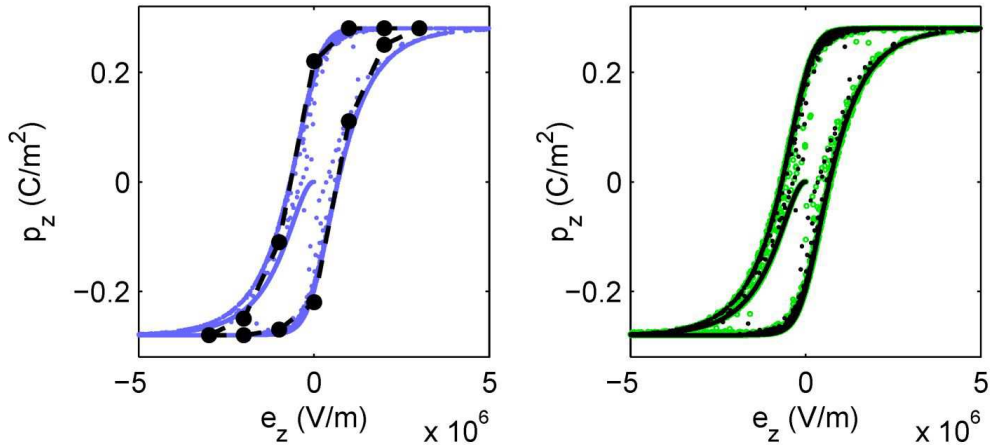}
\caption{These graphs display  histories of computed field components  located at the cavity point $(x,y,z)=(2.45$m$,2.45$m$,1.36$m$)$ (which is inside the rotating ferroelectric cylinder). For small amplitude drive currents in the cavity hysteretic polarization loci will exhibit different features at different locations in the ferroelectric medium.
For sufficiently large amplitude drive currents all such loci can be driven to limit cycles that exhibit saturation.
In these computations the model parameters do not vary with position in the medium so the resulting limit cycles are independent of position.
A typical ferroelectric hysteresis loop on the left displays the evolution of the $z-$component of electrical polarization inside the cylinder,  at rest at $z=1.36$m{\color{black}; small light (blue) circles indicate data points from the simulation.} This history is induced by the $z-$component of a steady $250$Hz electric field driven by the external (initially zero) current in the cavity. The evolution proceeds from the origin where ${\cal P}_z$ and ${\cal E}_z$ are zero to a time where a limit cycle is in evidence. The parameters of the model ($\alpha^{pe}=3.6 \times 10^{4}, \beta^{pe}=2.0 \times 10^{-6} $m/V$ , \xi^{pe}=1.3 \times 10^{5} $m$^{2}$/C$, \kappa^{pe}=0.5, \theta^{pe}=0.5, \sigma= 2.6 \times 10^{-4}$S/m) have been chosen so  that experimental data points (large black circles) from reference \cite{boch} lie close to this particular limit cycle.
In this simulation  the changes  in the  ${\cal P}_z$ vs ${\cal E}_z$ hysteresis loops induced by rotation  are  imperceptible  on the scales employed in the figure.
The light (green) points in the right graph result when the computation is repeated but with the cylinder rotating at $1497$rpm (for comparison dark (black) data points refer to the non-rotating case).
{\color{black}
%light (green) and dark (black) data points refer to the rotating and stationary cases respectively.
Each data point on each graph corresponds to a numerical solution sampled at equidistant time intervals.  The envelopes are produced by dense sets of data points while individual interior data points reflect transient configurations before the source current attains a purely sinusoidal time variation.}
}
\label{figure2}
\end{figure}

\begin{figure}[!h]
\centering
\includegraphics[scale=1.2]{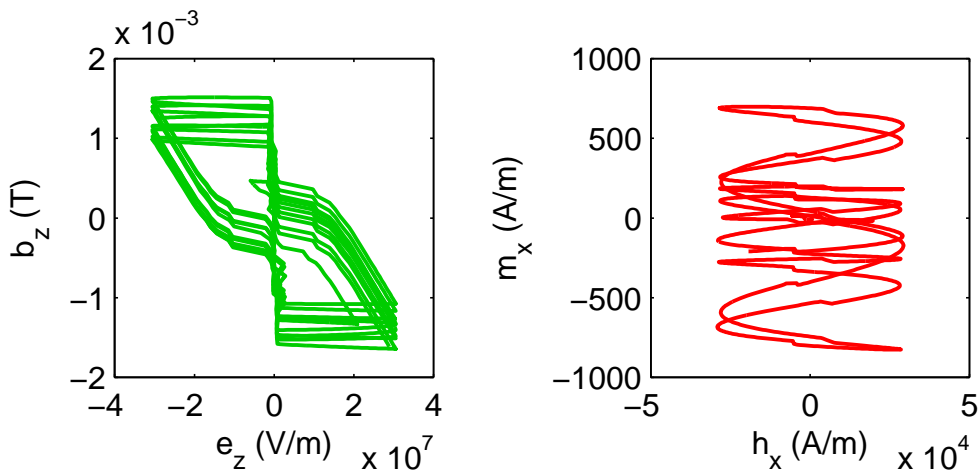}
\caption{ These graphs display  histories of computed field components  located at the same cavity point as in figure \ref{figure2} driven by a the external current outside the cylinder {\color{black} causing the cyclic behavior in the fields; there are about $10$ current source cycles, therefore about $10$ cycles per one revolution of the cylinder as can be seen on the left graph}. During the displayed history  the cylinder performs 1 complete revolution.
The  graph on the left displays the evolution of the $z-$components  ${\cal B}_z$ and ${\cal E}_z$, located at this cavity point,  projected onto a region of the ${\cal B}_z$-${\cal E}_z$ plane. {\color{black} It may be noted that in addition to the quasi-periodicity on the left graph the hysteretic envelope varies with time.}  On the right  a non-zero magnetization ${\cal M}_x$ induced by the  time-dependent ${\cal H}_x$ field  is displayed. Similar behavior is computed for  ${\cal M}_y$ as a function of the magnetic field and indicates the presence of a rotation induced  hysteretic magnetisation in the ferroelectric medium that increases in magnitude with the rotation speed $\Omega$.  {\color{black} The behavior of the curve on the left is correlated with the behavior of the curve on the right.  Thus when ${\cal B}_z$ (on the left) reaches saturation ${\cal H}_x$ (on the right) is at a local maxima or minima.}
}
\label{figure3}
\end{figure}

To analyze these equations numerically Yee's FDTD algorithm \cite{yee},\cite{tef}  for standard materials on a standard
staggered $E-H$ mesh has been extended to include non-linear hysteretic media with branched
solutions. Since $\nabla\cdot\mathbf{B}^U$ and $\nabla\cdot\mathbf{D}^U$ in the cavity are chosen
initially zero and the total external current is constructed to be divergenceless the full set of
material Maxwell equations is accommodated.
{\color{black}

Since the rotating cylinder occupies a finite region of space inside the vacuum cavity one is confronted with the problem of implementing electromagnetic interface conditions in an FDTD scheme.
A recent survey of computational algorithms that tackle this issue for both finite element and finite difference schemes can be found in \cite{cary}.
This reference also offers a more accurate finite difference algorithm than used in this paper.
However the approach adopted here has proved stable in the parameter domains explored and is based on the idea of `smoothing' the cylindrical interface with the vacuum region by modulating constitutive constants by a radial bump function that effectively approximates the characteristic domain occupied by the cylinder.
Such an approach, when digitized by using the Yee algorithm offers a good approximation to exact analytic results for static electric or magnetic fields outside and inside media at rest and is sufficiently accurate for our purposes when used to analyze Maxwell's equations for the rotating ferroelectric.

The details of each simulation are outlined in the figure captions \ref{figure2} and \ref{figure3}.  One may regard the solutions for the fields
$\{{\cal E}_{x}^{U}$, ${\cal E}_{y}^{U}$, ${\cal E}_{z}^{U}$,
${\cal P}_{z}^{U}$,
${\cal H}_{x}^{U}$, ${\cal H}_{y}^{U}$, ${\cal H}_{z}^{U}$,
${\cal M}_{x}^{U}$, ${\cal M}_{y}^{U}\}$ at each point $(x,y,z)$ in the rotating cylinder as a space curve parameterized by time $t$ in $9$ dimensions.  Such curves will not in general be closed curves in 9 dimensions unless the rotation frequency and the frequency of the harmonic steady source current are rationally related in the same units.  These curves will not in general have closed projections on any particular $2$ dimensional plane (e.g. the ${\cal M}_{x}^{U}-{\cal H}_{x}^{U}$ plane).
Although the projections in figure \ref{figure2} display an imperceptible change when the cylinder rotates at zero and $1497$rpm the projections in figure \ref{figure3}, by contrast, indicate a significant change at $1497$rpm.

}

\section{Conclusions}
A class of non-linear constitutive relations for materials with memory has been discussed  in the framework of covariant macroscopic Maxwell theory.
The general approach enables models to be formulated for arbitrarily moving media including those that exhibit hysteretic responses to
time varying electromagnetic fields.
  Using a particular parameterized model, consistent with experimental data for a particular material that exhibits purely ferroelectric hysteresis when
at rest in a slowly varying electric field,  a  numerical  analysis of its response to a driven harmonic electromagnetic field in a rectangular cavity has
been performed when in different states of rotation about its `soft' direction. The results indicate that such a model offers a means to compute numerically
the significance of induced hysteretic magnetisation in a ferroelectric medium as a function of its rotation speed and the frequency of an external self-consistent electromagnetic field.

\section{Acknowledgements}

The authors are members of the ALPHA-X collaboration funded by EPSRC and are also grateful for support from the Cockcroft Institute
of Accelerator Science and Technology (STFC).

\end{document}